\documentclass[a4paper, twocolumn]{article}

\usepackage{amsmath}
\usepackage{amsfonts}
\usepackage{graphicx}
\usepackage[colorlinks=true, allcolors=blue]{hyperref}
\usepackage{mathrsfs}
\usepackage{amssymb}
\usepackage{commath}
\usepackage[sorting=none,style=numeric]{biblatex}
\usepackage{nccmath}
\usepackage{lipsum}
\usepackage{dblfloatfix}

\renewcommand{\vec}[1]{\mathbf{#1}}
\newcommand*\dyad[1]{\overset{\text{\tiny$\leftrightarrow$}}{#1}}
\newcommand\rmi[1][]{\mathrm{i}}
\newcommand\rmd[1][]{\mathrm{d}}
\newcommand\rme[1][]{\mathrm{e}}
\bibliography{acs-achemso}

\usepackage{geometry}
\begin{document}
\twocolumn[
   \begin{center}
     {\huge\sffamily Dispersion-induced \(Q\)-factor enhancement in waveguide-coupled surface lattice resonances}\\
      \vspace{2ex}
      \textsc{Jussi Kelavuori, Ali Panah Pour, and Mikko J. Huttunen}\\
      \vspace{2ex}
      \textit{Photonics Laboratory, Physics Unit, Tampere University, FI-33014 Tampere, Finland}
   \end{center}
   ]

\begin{abstract}
Diffractively coupled nanoparticle arrays are promising candidates for helping to flatten many photonic devices such as lasers, lenses, and metrology instruments. Their performance, however, is directly linked with the size of the metasurfaces, limiting their applicability in nanophotonic applications. Here, we dramatically reduce array sizes of high-\(Q\)-factor metasurfaces by utilizing strongly dispersive media. The effect is demonstrated by theoretically and numerically studying periodic arrays of plasmonic nanoparticles embedded inside Bragg-reflector waveguides. We demonstrate array dimensions reduction up to two orders of magnitude while still achieving ultra-high \(Q\)-factors in excess of 10\(^4\).
\end{abstract}

\section{Introduction}
Plasmonics, a subfield of photonics, utilizes the properties of interfaces between metals and dielectric materials to couple electromagnetic radiation with charge currents in the metals. Fabrication methods such as electron-beam and ion-beam lithographies~\cite{Horak2018} allow nanometer-scale control over the plasmonic structures, enabling the realization of metamaterials with exotic optical properties. Arrays of periodically placed nanoparticles (NPs) with interparticle distances close to the incident wavelength support so-called surface lattice resonances (SLRs), which couple the diffractive orders of the array with the individual plasmonic responses of the NPs. Plasmonic--diffractive hybrid resonances are the basis for many technologies demonstrated in plasmonic metamaterials, such as lasing~\cite{Zhou2013,Azzam2020}, nonlinear optics~\cite{Stolt2022}, and sensing~\cite{Danilov2018, Magno2018}. The usefulness of SLRs stems from the strong light--matter interaction typical for plasmonic resonances, while simultaneously achieving significantly decreased radiative losses, leading to even stronger local electric fields near the particles. 

The ability of a resonator to store energy is quantified using the quality factor (\(Q\)-factor), 
relating the energy loss per resonance cycle to overall resonance energy. Due to diffractive coupling, SLRs have naturally high \(Q\)-factors compared to other plasmonic resonances, with \(Q\)-factors higher than 2\,000 observed~\cite{Bin-Alam2021}. The inherent ohmic losses in the plasmonic NPs, however, limit the \(Q\)-factors to inferior values compared to similar dielectric resonances such as quasi-bound states in the continuum~\cite{Liu2019, Jin2019}.

Techniques to control and increase \(Q\)-factors in SLRs focus on controlling either the absorptive or radiative losses. The plasmonic absorptive losses are usually decreased by increasing the resonance wavelength. Consequently, the highest \(Q\)-factors in SLRs are typically achieved in the mid-infrared region~\cite{Bin-Alam2021, Li2014}. Remarkably, recent theoretical work  suggests that the absorptive losses in plasmonic structures can be avoided completely by cleverly engineering coupling between three different optical modes~\cite{Kolkowski2023}. Conversely, the radiative losses in SLRs can be decreased by increasing either the array size~\cite{Bin-Alam2021,Zundel_2019} or the light source spatial coherence~\cite{Bin-Alam2021}. Furthermore, utilizing out-of-plane resonances~\cite{Li2014} and quadrupolar coupling \cite{Fang:21}
might increase \(Q\)-factors. Moreover, the symmetry of the dielectric surroundings of the lattice has been shown to affect the \(Q\)-factor drastically, allowing active control over the resonance linewidth~\cite{Kelavuori2021}. In optical microcavities, strong dispersion in the medium has been utilized to increase the \(Q\)-factors of the cavities~\cite{Soljacic2005, Gao2016}. Slow-light effects arising from strong material dispersion also allow decreasing the sizes of some nanophotonic devices, such as optical switches~\cite{Vlasov2005,Beggs:08,Baba2008}.

The radiative losses in SLRs are determined by how well the associated mode is coupled with the far field~\cite{Kuhner2023}. This coupling is weaker for larger arrays or increased light source coherence, leading to enhanced \(Q\)-factors. Large arrays are very effective at prohibiting radiative decay, with radiant \(Q\)-factors of 5\,000 reached in large arrays~\cite{Li2014}.  On the other hand, bound states in the continuum are completely uncoupled from the far field, yielding them with infinite radiant \(Q\)-factors. Correspondingly, coupling into such modes is impossible from the far field. While increasing array size decreases radiative losses effectively, the leaking radiation can also be stopped by constructing the surroundings accordingly. The effect can be achieved in, for example, microring coupled NPs~\cite{Chamanzar:11}. Moreover, coupling the diffractive orders through waveguide modes is also advantageous since the scattered fields are more confined in the lattice plane. These waveguide-mode coupled plasmonic--diffractive resonances have been called waveguide plasmon polaritons~\cite{Christ2003,Rodriguez2012}, guided lattice resonances~\cite{Abir2022} and waveguided plasmonic surface lattice resonances~\cite{Ugulen2022}. Still, their advantage over free-space coupled SLRs is limited since traditional dielectric waveguides trap only a limited angular range.

Here, we numerically investigate the formation and characteristics of SLRs, when NPs are placed inside strongly dispersive media. A simple and strongly dispersive system is realized by embedding plasmonic NP arrays inside Bragg reflector waveguides (BRWs), operated close to their cut-off region. The properties of SLRs are calculated by extending the conventional lattice-sum approach (LSA) formulation \cite{Huttunen2016} to the studied mirror waveguide system. The radiant \(Q\)-factors are shown to be inversely proportional to the group index of refraction at the surroundings of the NP array. The strong dispersion of BRWs near their cut-off region significantly increases the radiant \(Q\)-factor, reaching values close to 12,000 with NP lattices of only 50 particles. The computational results demonstrate the effectiveness of using highly dispersive materials for achieving small-area--ultra-high-Q plasmonic structures. Such metasurfaces would be a step towards industrial plasmonic metasurfaces due to significantly decreased fabrication write time and device footprint. Furthermore, the reduced metasurface dimensions pave the way for utilizing SLR arrays as pixels in, for example, spatial light modulators or hyperspectral imaging applications.

\section{Theory}{\label{ch:theory}}
Periodic NP arrays can support hybrid plasmonic--diffractive resonances known as SLRs. Since most of the losses associated with SLRs are associated with the plasmon oscillations, high-\(Q\) SLRs are usually designed to be spectrally far away from the single-particle plasmonic resonance. This also makes their wavelength almost fully dictated by the diffractive mode of the system, with only a minor shift arising from the phase delays related to the scattering in the individual NPs. For an array of scatterers, the resonance energy of the diffractive mode is dictated by the momentum conservation equation for grating coupling \cite{Liberman:11}:
\begin{equation}{\label{eq:slrloc}}
    k_{\mathrm{sub}} = k_\mathrm{inc}+mk_g\,,
\end{equation}
where \(k_\mathrm{sub}\) is the wavenumber in the substrate, \(k_\mathrm{inc} =\sin{(\theta)}k_0\) is the tangential component of the incident wavevector, \(\theta\) is the incident angle, \(k_0 = 2\pi/\lambda\) is the free-space wavenumber, \(m\in \mathbb{Z}\) is the diffractive order, and  \(k_g = 2\pi/p\) is the grating wavenumber with an array period \(p\). 

In free-space SLRs, the substrate wavenumber \(k_\mathrm{sub}\) is affected by the bulk refractive index \(n_\mathrm{sub}\) while in guided-mode SLRs the waveguide-mode specific effective refractive index \(n_\mathrm{eff}\) dictates the resonance condition with \(k_\mathrm{sub} = n_\mathrm{eff}k_0\). In traditional dielectric waveguides, this does not substantially modify the resonance properties from their free-space counterparts, since the effective index is limited to values between the used bulk materials. In mirror waveguides, however, the effective index approaches zero at the so-called cut-off frequency of the waveguide mode. Designing an SLR in this region would imply exotic resonance conditions, where the resonance wavelength would be more strongly dictated by the angle of incidence \(\theta\).

\begin{figure*}
    \centering
    \includegraphics{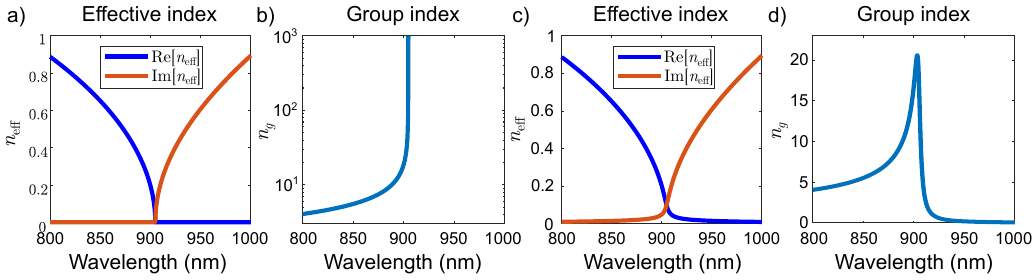}
    \caption{Dispersion of effective refractive indices for mirror waveguides. (a) Effective phase and (b) group indices for a perfect-mirror waveguide system. (c) Effective phase and (d) group indices for a lossy-mirror waveguide system with \(k_c^{\prime\prime}=40\)\;mm\(^{-1}\).}
    \label{fig:dispcurves}
\end{figure*}

The effective refractive index in planar perfect-mirror waveguides is given by the following dispersion relation:
\begin{equation}
    n_\mathrm{eff} = \frac{k_z}{k_0} = \sqrt{n_\mathrm{core}-\frac{k_c^2}{k_0^2}}\,,
\end{equation}
where \(k_z\) is the wavenumber in the propagation direction,  \(n_\mathrm{core}\) is the core bulk refractive index, \(k_c = \frac{\mathrm{m}\pi}{b}\) is the cut-off, i.e.~transverse wavenumber for the waveguide mode of order \(\mathrm{m}\), and \(b\) is the height of the waveguide. As shown in Fig.~\ref{fig:dispcurves} the effective index in perfect-mirror waveguides approaches zero at the cut-off wavelength.

Interestingly, the group index
\begin{equation}
    n_g = \frac{c}{v_g} = n_\mathrm{eff} + \omega\frac{\rmd n_\mathrm{eff}}{\rmd \omega}
\end{equation}
approaches infinity at the cut-off wavelength in perfect-mirror waveguides, due to the extreme dispersion near the cut-off. This behavior is only possible due to an idealized model of perfectly reflecting surfaces. In physical mirror waveguides, such as metal- and Bragg-mirror waveguides, the inherent material or radiative losses limit the achievable group indices to finite values. To approximate the lossy-mirror-waveguide dispersion relations, we introduce an imaginary part to the cut-off wavenumber \(k_c=k_c^{\prime}+\rmi k_c^{\prime \prime}\), where \(k_c^{\prime}, k_c^{\prime \prime} \in \mathbb{R}\). The modification leads to lossy-mirror waveguide dispersion relations~\cite{Hu:09}:
\begin{equation}{\label{eq:kz_lossy}}
    k_z = n_\mathrm{eff}k_0 =  \sqrt{k_0^2n_\mathrm{core}^2-k_c^{\prime 2}-2\rmi k_c^{\prime} k_c^{\prime\prime}+k_c^{\prime\prime 2}}\,,
\end{equation}
where the complex cut-off wavenumber \(k_c\) is assumed constant for a given waveguide mode, similar to fully guided waveguide modes. The effective and group indices are shown in Figs.~\ref{fig:dispcurves}(c) and (d), respectively, for a lossy-mirror waveguide with \(k_c^{\prime\prime}=40\)\;mm\(^{-1}\). The losses in the system limit the dispersion at the cut-off wavelength to a set value limiting the maximum group index for the waveguide. 

In this work, we deploy the lattice-sum approach (LSA) to estimate the properties of mirror-waveguide-coupled SLRs. The method uses dipole approximation, making each particles response dictated only by its dipole moment \(\vec{p}=\varepsilon_0 \dyad{\alpha}\vec{E}\), where \(\vec{E}\) is the incident electric field, \(\dyad{\alpha}\) is the particle polarizability and \(\varepsilon_0\) is the permittivity of free space. In LSA specifically, each NP of the nano-array is assumed to have identical dipole moment. Using these assumptions, the collective response of the particles can be written using the effective polarizability \(\dyad{\alpha}_\mathrm{eff}\) \cite{Huttunen2016}:
\begin{equation}\label{eq:alphaeff}
    \dyad{\alpha}_\mathrm{eff} = \left(\dyad{\alpha}^{-1}-\varepsilon_0\dyad{S}\right)^{-1}\,,
\end{equation}
where \(\dyad{\alpha}\) is the single particle polarizability and \(\dyad{S}\) is the lattice sum of the array. The effective polarizability \(\dyad{\alpha}_\mathrm{eff}\) can be used to calculate the extinction cross-section of the metasurface~\cite{Huttunen2016}:
\begin{equation}
    \sigma_\mathrm{ext} = \frac{k_0}{A}\mathrm{Im}[\alpha_\mathrm{eff}]\,,
\end{equation}
where \(A\) is the unit cell area in the array.

While single-particle polarizabilities are in general second-order tensors, the considered spherical nanoparticles have isotropic, i.e.~scalar polarizabilities. Consequently, the polarizability $\alpha$ can be determined using the first order (dipole) approximation in Mie-theory~\cite{Doyle1989}:
\begin{equation}{\label{eq:Mie}}
    \alpha = \rmi \frac{3l^3}{2x^3}\frac{m\psi_1(mx)\psi_1^{\prime}(x)-\psi_1(x)\psi_1^{\prime}(mx)}{\psi_1(mx)\xi_1^{\prime}(x)-m\xi_1(x)\psi_1^{\prime}(mx)}\,,
\end{equation}
where \(l\) is the radius of the sphere, \(x = 2\pi n_h l/\lambda\) is the size factor, \(m=n_s/n_h\) is the ratio between the refractive indices of the sphere $n_s$ and the host material $n_h$, while \(\psi_1\) and \(\xi_1\) are the first-order Ricatti-Bessel functions of first and second kind, respectively. 

The lattice sum \(\dyad{S}\), on the other hand,  describes the collective effect of the array on the induced dipole moment through scattered fields. It is written as
\begin{equation}{\label{eq:latticesum}}
    \dyad{S} = \omega^2\mu_0\sum_{j\neq i}^N \dyad{G}_e(\vec{r}_i,\vec{r}_j)\,.
    \end{equation}
Here, \(\omega\) is the angular frequency of the incident light, \(\mu_0\) is the vacuum permeability, \(N\) is the total number of particles in the array and \(\dyad{G}_e\) is the dyadic electric Green's function from  \(\vec{r}_j\) to \(\vec{r}_i\). In general, the electric Green's function can be used to calculate the electric field an electric dipole induces to its surroundings:
\begin{equation}
    \vec{E}(\vec{r}_i) = \omega^2\mu_0\dyad{G}_e(\vec{r}_i, \vec{r}_j)\vec{p}(\vec{r}_j)\,.
\end{equation}
Therefore the lattice sum can be understood as a sum over the scattered electric fields from an array of particles at a one central particle \(\vec{r}_i\) in an empty lattice, where the lattice sites have not yet been specified.

Conventional LSA formulations are based on the free-space electric Green's function. However, this formulation is only applicable when the radiation pattern of the scatterers is not disrupted by any major interfaces in the surrounding geometry. In waveguides, for example, some of the scattered light is radiated into the waveguide modes, a phenomenon not described by the free-space Green's function. In practice, electric Green's functions are easily obtained only in a handful of different geometries. An example of a relatively simple geometry is the rectangular perfect electric conductor (PEC) waveguide consisting of a hollow core with walls of lossless mirrors. Its dyadic electric Green's function is given by~\cite{Tai1972}: 
\begin{equation}{\label{eq:Ge1}}
\begin{aligned}
\dyad{G}_{e 1}\left(\vec{R}, \vec{R}^{\prime}\right)
= & -\frac{1}{k^2} \hat{z} \hat{z} \delta\left(\vec{R}-\vec{R}^{\prime}\right) \\
& +\frac{\rmi}{a b} \sum_{\mathrm{m,n}} \frac{2-\delta_0}{k_c^2 k_z}\left[\vec{M}_{\mathrm{mn}}\left(\pm k_z\right) \vec{M}_{\mathrm{mn}}^{\prime}\left(\mp k_z\right)\right. \\
& \left.+\vec{N}_{\mathrm{mn}}\left(\pm k_z\right) \vec{N}_{\mathrm{mn}}^{\prime}\left(\mp k_z\right)\right], z \gtrless
z^{\prime} .
\end{aligned}
\end{equation}
The primed variables refer to dipole source coordinates and the unprimed variables to field evaluation point. Hat variable denotes a unit vector and \(\delta_0 = 1\) if \(m=0 ~\;\lor~\; n = 0\) and \(\delta_0 = 0\) otherwise. The sign of the arguments for the vector wave functions \(\vec{M}_{\mathrm{mn}}\) and \(\vec{N}_{\mathrm{mn}}\) is determined by whether the source is behind or in front of the evaluation point (\( z \gtrless
z^{\prime}\)). The vector wave functions were determined using Dirichlet boundary conditions for a PEC waveguide with width (\(x\)) of \(a\) and height (\(y\)) of \(b\). The wave functions can be written as \cite{Tai1972}
\begin{equation}{\label{eq:wavefuncs}}
\begin{aligned}
\vec{M}_{\mathrm{mn}}(k_z)=&\left(-k_y\cos(k_xx)\sin(k_yy)\hat{x}+ \right. \\
& \left. k_x\sin(k_xx)\cos(k_yy)\hat{y}\right)\rme^{\rmi k_zz}, \\
\vec{N}_{\mathrm{mn}}(k_z)= &\frac{1}{k}\left( \rmi k_zk_x\cos(k_xx)\sin(k_yy)\hat{x}+\right. \\ 
&\left. \rmi k_zk_y\sin(k_xx)\cos(k_yy)\hat{y}+\right.\\
&\left. k^2_c\sin(k_xx)\sin(k_yy)\hat{z}\right)\rme^{\rmi k_zz}
\end{aligned}
\end{equation}
with wavenumbers specified as:
\begin{equation}{\label{eq:greenhelp}}
\begin{aligned}
 k_x &= \left(\frac{\mathrm{m}\pi}{a} \right), ~  k_y = \left(\frac{\mathrm{n}\pi}{b} \right) \\
 k^2 &= k_x^2+k_y^2+k_z^2 = k_c^2+k_z^2\,.
\end{aligned}
\end{equation}
While the PEC waveguide is an interesting academic example, practical systems utilizing e.g.~metallic mirrors are associated with non-negligible losses. For a more realistic Green's function, we replace the perfect mirror propagation wavenumber \(k_z\) from Eqs.~\eqref{eq:greenhelp} with the lossy-mirror waveguide wavenumber from Eq.~\eqref{eq:kz_lossy}. In practice, this modification introduces either exponential loss or gain in the propagation direction \(z\) depending on the sign of the imaginary part of the transverse wavenumber \(k_c\).

Applying the rectangular lossy-mirror waveguide Green's function to the lattice sum described in Eq.~\eqref{eq:latticesum}, we can now calculate how an array of scatterers behaves when embedded inside a mirror waveguide. Specifically, we focus on a 1D array periodic in the propagation \(z\) direction, with each particle having identical \(x\) and \(y\) coordinates. By noting that the Green's function is dependent on the \(z\)-coordinate only through the phase term \(\rme^{\rmi k_zz}\), the lattice sum \(\dyad{S}\) of this array can be expressed as a geometric sum over the particle locations. For an infinite number of particles, the sum converges to a geometric series, and the lattice sum can be expressed in a closed form: 
\begin{equation}{\label{eq:closedformeq}}
\begin{aligned}
     \dyad{S} = -\omega^2 \mu \sum_{\mathrm{mn}}\dyad{G}_{\mathrm{mn}} & \left( \frac{1}{1-\rme^{-\rmi p(k_z+k_{\mathrm{inc},z})}}+\right.\\
     &\left. \frac{1}{1-\rme^{-\rmi p(k_z-k_{\mathrm{inc},z})}}-1 \right)\,,\\     
\end{aligned}
\end{equation}
where
\begin{equation}
\begin{aligned}
\dyad{G}_{\mathrm{mn}} = &\frac{2-\delta_0}{k_c^2 k_z}\left(\vec{M}_{\mathrm{mn}}^* \vec{M}_{\mathrm{mn}}^{*\prime}+\vec{N}_{\mathrm{mn}}^*\vec{N}_{\mathrm{mn}}^{*\prime}\right)\,,\\ 
 \end{aligned}
\end{equation}
and the modified vector wave functions \(\vec{M}_{\mathrm{mn}}^* \), \( \vec{M}_{\mathrm{mn}}^{*\prime}\), \( \vec{N}_{\mathrm{mn}}^*\) and \(\vec{N}_{\mathrm{mn}}\) are devoid of the forward-propagation phase term \(\rme^{\rmi k_zz}\). Here, \(p\) is the array period and \(k_{\mathrm{inc},z}\) is the \(z\)-component of the incident wavevector.  For detailed derivations of the lattice sum for both the mono- and multi-partite unit cells, refer to Supplemental material [\textbf{LINK HERE BY THE PUBLISHER}]. Note that the wavenumbers (\(k_x\), \(k_y\), \(k_z\)) differ for each waveguide mode (\(\mathrm{m}\), \(\mathrm{n}\)) leading to different coupling conditions for each mode. Embedded scatterers may therefore be used as mode selective waveguide grating couplers, with a possibility to suppress diffractive coupling for selected waveguide modes (see Supplemental material [\textbf{LINK HERE BY THE PUBLISHER}]).

The used approach does not take into account the uncertainty associated with the leaky waveguide mode wavenumbers. While the losses over propagated distance are included in the model, a \(\delta\)-function-like effective index for each frequency is assumed. For accuracy, a Lorentzian distribution of propagation wavenumbers \(k_z\) for each wavelength should be assumed for leaky waveguide modes, with the accompanied linewidth proportional to the imaginary part of the effective index~\cite{Hu:09}. Applied to SLRs, the effect broadens resonance conditions and slightly reduces \(Q\)-factors, especially near the cut-off frequency where \(\mathrm{Im}[n_\mathrm{eff}]\) is at its highest. The effect is discussed more carefully in Supplemental material [\textbf{LINK HERE BY THE PUBLISHER}].
\section{Results and discussion}

\begin{figure*}[btp]
    \centering
    \includegraphics[scale=1]{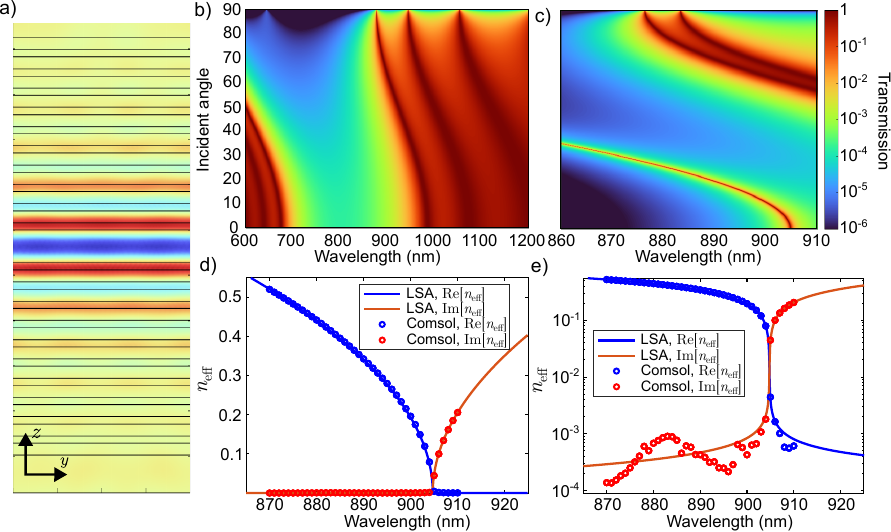}
    \caption{(a) TE\(_0\) planar waveguide mode field profile for \(E_y\). Transmission spectra of the (b) used Bragg mirror, and (c) the full BRW. (d),(e) Complex dispersion curves for the TE\(_0\)-planar waveguide mode in COMSOL, and the parameter-fitted lossy-mirror waveguide dispersion curve for the rectangular TE\(_{10}\)-mode .}
    \label{fig:WGprop}
\end{figure*}
To demonstrate the potential of mirror waveguides as a medium for ultra-high-\(Q\) SLRs, we employ the discussed theoretical framework for BRWs.  The main advantage of using BRWs, instead of metal waveguides, concerns their losses at optical and near-infra red frequencies. BRWs support waveguide modes with close to zero effective refractive indices. In addition, group indices as high as \(n_g = 40\) have been demonstrated earlier~\cite{Fuchida2012}. The investigated BRW consists of a planar waveguiding layer surrounded by ten pairs of quarter-wave thick high- and low-index materials on both sides of the waveguide. TiO\(_2\)/BK7-glass was used as a high/low index material, with the waveguiding layer made of BK7. Layer thicknesses are presented in Table \ref{table:braggtech}. 

\begin{table}[b!]
\centering
\begin{tabular}{|l|l|l|ll}
\cline{1-3}
Layer          & Material & Thickness   &  &  \\ \cline{1-3}
High index QWP & TiO\(_2\)   & \(80.1274\)\;nm  &  &  \\ \cline{1-3}
Low index QWP  & BK7      & \(132.5688\)\;nm &  &  \\ \cline{1-3}
Guiding layer  & BK7      & \(360\)\;nm      &  &  \\ \cline{1-3}
\end{tabular}
\caption{Layer data used in COMSOL simulations.}
\label{table:braggtech}
\end{table}

To couple electromagnetic radiation with the NPs embedded inside the BRW, the angle of incidence was carefully considered. Given the angle-dependent reflectance of Bragg reflectors, depicted in Fig.~\ref{fig:WGprop}(b), the system was designed to be transparent for highly oblique angles near the wavelength 900\;nm. Consequently, incident radiation can efficiently couple with the plasmonic scatterers inside the waveguide. Since the incident wavelength is close to a cut-off wavelength of the TE\(_0\)-mode of the waveguide, the NPs will scatter light dominantly to the waveguide mode due to its enhanced Purcell factor. Since the effective index of the mode is close to zero, the mode wave vector is directed close to the normal of the Bragg walls. High reflectivity at normal incidence results in the TE\(_0\)-mode having low losses in the used wavelength range.
To fully take advantage of this effect, the BRW-guided SLR can be engineered to appear close to 900\;nm range by utilizing Eq.~\eqref{eq:slrloc}, and modifying the lattice period accordingly. 

At cut-off, the produced mode can be alternatively understood as the plasmonic particle resonance hybridizing with the epsilon-near-zero (ENZ) mode of the thin-film structure~\cite{GERASIMOV2019}. While the plasmonic contribution negatively affects the \(Q\)-factor of the pure ENZ-mode, the small mode volumes obtainable in the NPs grant them advantages over all-dielectric structures~\cite{Bozhevolnyi:16}. The reflection of the structure without the NPs is shown in Fig.~\ref{fig:WGprop}(c), where the ENZ/Bragg-wall-cavity mode can be seen as a thin line starting from 905\;nm at normal incidence. While the BRW structure restrains the electric fields in the direction normal to the Bragg walls, the periodic array imposes phase restrictions in the direction of the lattice vector. Furthermore, field confinement in the NPs results in higher local fields and increased applicability.

Modal analysis was done for the described BRW with the finite-element COMSOL Multiphysics program. The dispersion graphs for the real and imaginary parts of the effective index of the TE\(_0\)-mode are shown in Figs. \ref{fig:WGprop}(d) and (e) with linear and logarithmic scaling, respectively. The lossy-mirror waveguide dispersion relation given by Eq.~\eqref{eq:kz_lossy} for the (rectangular) TE\(_{10}\)-mode was then fitted to the obtained data to be used for LSA. The field profile of the rectangular mirror waveguide mode  TE\(_{10}\) resembles the planar waveguide TE\(_0\) field profile with both varying only in the vertical (\(y\)) direction. The fitted values are \(n_\mathrm{core}=1.9\), waveguiding layer thickness \(b=238.1\)\;nm an  \(\mathrm{Im}[k_c] = k_c^{\prime\prime} = 600\)\;m\(^{-1}\). The fitted complex effective index of a lossy-mirror waveguide mode is shown in Figs.~\ref{fig:WGprop}(d) and (e), with a good correspondence to the COMSOL simulations. Variations in the imaginary part of \(n_\mathrm{eff}\) in COMSOL simulations are due to reflections from the finite-sized perfectly-matched layers. The proper imaginary part of the mode is found between the extrema of these oscillations.~\cite{Hu:09}

\begin{figure}[btp]
    \centering
    \includegraphics{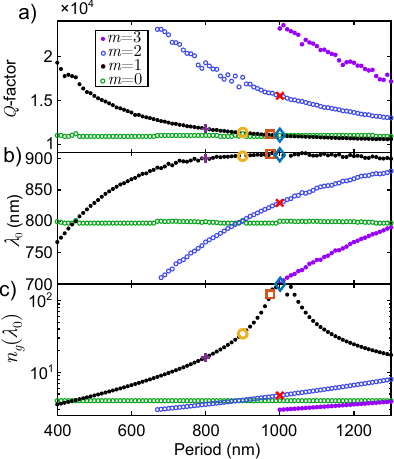}
    \caption{Properties of SLRs coupled via the fitted lossy-mirror TE\(_{01}\)-mode. (a) Lattice \(Q\)-factors (b) resonance wavelengths, and (c) TE\(_{01}\)-mode group indices at the resonance wavelengths for infinite arrays found within the wavelength range \(700-905\)\;nm as a function of lattice period.}
    \label{fig:Infsum}
\end{figure}
\begin{figure*}[btp]
    \centering
    \includegraphics[]{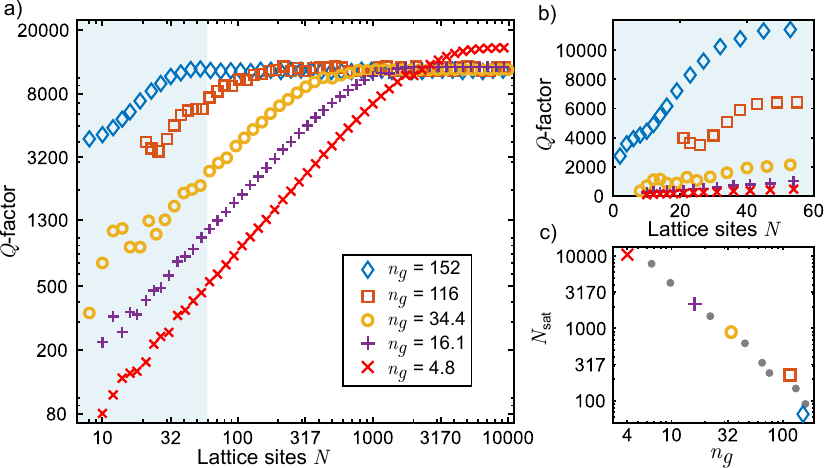}
    \caption{ \(Q\)-factors of selected lattice sums as a function of total particles in the system in a (a) logarithmic and (b) linear scale. (c) Saturation particle count as a function of mode group index.}
    \label{fig:Saturation}
\end{figure*}
LSA was then applied to study the lossy-mirror-waveguide system. The lattice sites for NPs were placed in the middle of the waveguide, and only the response from the TE\(_{10}\) mode was considered. The used incident angle for the system was 64 degrees, aligning in the low-reflection region of the Bragg-reflector in Fig.~\ref{fig:WGprop}(b) at wavelengths close to \(900\)\;nm. The results from LSA, namely the lattice sum and transmission, were analyzed by fitting a Fano function to the data~\cite{Wang:19}: 
\begin{equation}{\label{eq:fano}}
    T_{\mathrm{fano}}(\lambda) = |a_1+a_2\rmi+\frac{I}{(\lambda-\lambda_0-\rmi\gamma)}|^2\,,
\end{equation}
where \(\lambda_0\) is the resonance wavelength, \(\gamma\) is the resonance half-width half-maximum, and \(a_1,~a_2\) and \(I\) are other fitting parameters determining the shape and strength of the resonance. The resonance \(Q\)-factor is given as \(Q=\frac{\lambda_0}{2\gamma}\) from the fitted function.  Some examples of data-fitted functions are presented in Supplemental material [\textbf{LINK HERE BY THE PUBLISHER}].

We begin the analysis with the empty-lattice approximation (ELA) of our system by only considering lattice sites devoid of any NPs.  This analysis of the lattice sum \(S\) is equivalent to a system with vanishingly small NPs. First, we analyze the ELA for an infinite array using Eq.~\eqref{eq:closedformeq}. Fig.~\ref{fig:Infsum} shows the variations in the ELA \(Q\)-factor, wavelength \(\lambda_0\), and the corresponding group index \(n_g(\lambda_0)\) as a function of the 1D array period \(p\). The resonance characteristics change continuously and quite predictably as a function of the lattice period for different SLR orders \(m\), with \(\lambda_0\) following closely the prediction from Eq. \eqref{eq:slrloc}. The symbolically highlighted cases are further analyzed in Fig.~\ref{fig:Saturation}.

With increasing period, the first-order coupling mode (\(m=1\)) approaches the cut-off wavelength of the waveguide mode at around 905\;nm. The cut-off is reached with a period of around 1\;\textmu m leading to a substantial increase in the group index. Intriguingly, the elevated group index does not affect the \(Q\)-factor. This phenomenon likely arises from two opposing factors counterbalancing each other as the cutoff is approached. While the elevated group index enhances the \(Q\)-factor~\cite{Soljacic2005,Gao2016}, the simultaneously increased losses (\(\mathrm{Im}[n_\mathrm{eff}]\)) reduce the coupling between far-away NPs consequently decreasing the \(Q\)-factor. Zeroth-order \(m=0\) resonances are analyzed more carefully in Supplemental material [\textbf{LINK HERE BY THE PUBLISHER}].

\begin{figure*}[btp]
    \centering
    \includegraphics[]{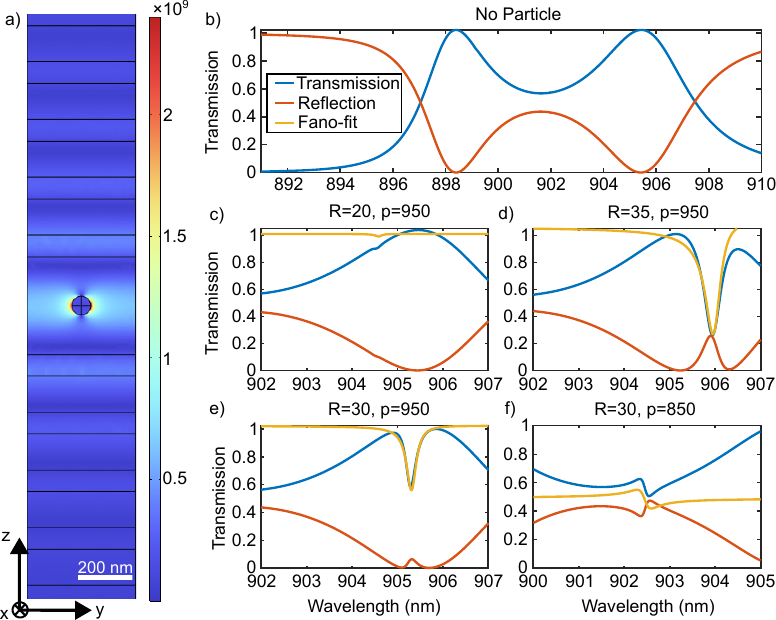}
    \caption{(a) Magnitude of \(E_y\)-field in the 3D COMSOL simulation at SLR resonance wavelength. (b) Transmission and reflection of the structure with no NPs. (c)--(f) Transmission, reflection and fitted Fano-function for BRW systems with spherical gold NPs of varying radii and array periods.}
    \label{fig:comsol}
\end{figure*}
While the group index hardly affects the \(Q\)-factor of infinite arrays, the underlying functions suggest finite arrays benefiting substantially from high-group-index coupling. Fig.~\ref{fig:Saturation} illustrates the ELA \(Q\)-factors with various group indices as a function of the total number of lattice sites, depicted in (a) log-log scale and (b) linear scale. The different group indices were obtained by varying the array periods and SLR orders as illustrated in Fig.~\ref{fig:Infsum}. For all cases, increasing lattice sites initially elevates the $Q$-factor until a saturation point is reached. Strikingly, this saturation occurs with significantly fewer lattice sites in regions with high group indices. For small arrays, it is evident that orders of magnitude \(Q\)-factor enhancement is achievable by utilizing the effect.

Further analyzing the effect of group index on array size, we calculated the saturation point for distinct arrays. The saturated lattice site number \(N_\mathrm{sat}\) was defined as the number of lattice sites required to achieve 98\% of the infinite array \(Q\)-factor. Our findings, depicted in Fig.~\ref{fig:Saturation}(c), indicate an almost inverse proportionality between the saturated lattice sites and the group index with a fit on the data suggesting proportionality of \(N_\mathrm{sat} \propto n_g^{-1.3}\). Before saturation, the addition of particles leads to a nearly linear growth in the \(Q\)-factor. Consequently, increasing the group index inversely impacts the number of particles necessary to attain specific \(Q\)-factors, serving as a useful rule of thumb.

\begin{figure}[btp]
    \centering
    \includegraphics[scale=1]{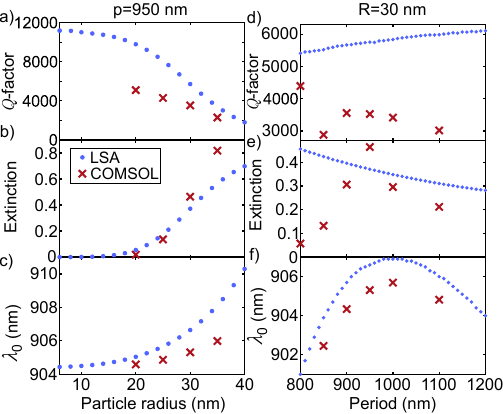}
    \caption{Properties of BRW-guided SLRs in LSA and COMSOl. (a),(d) \(Q\)-factors, (b),(e) maximum extinction, and (c),(f) resonance wavelength \(\lambda_0\) as a function NP radii and array period. }
    \label{fig:QwithAuCOMSOL}
\end{figure}

Since the findings presented in Figs.~\ref{fig:Infsum} and \ref{fig:Saturation} originate directly from ELA, the depicted \(Q\)-factors are related to the extent the metasurface geometry can mitigate radiative losses of the NPs. Considering, that the overall losses encompass both absorptive and radiative components according to the equation \(Q_\mathrm{tot}^{-1}=Q_\mathrm{abs}^{-1}+Q_\mathrm{rad}^{-1}\)~\cite{Li2014}, the relatively high radiant \(Q\)-factors do not necessarily translate to high total \(Q\)-factors for SLRs. To investigate the role of absorptive losses, different-sized gold NPs were introduced into the LSA simulation. The single-particle polarizabilities were determined using Eq.~\eqref{eq:Mie} using tabulated values for the permittivity of gold~\cite{Johnson1972}. The properties of first-order (\(m=1\)) infinite-array SLRs with different particle radii and array periods are depicted in Fig.~\ref{fig:QwithAuCOMSOL}. Increasing particle size leads to both higher scattering cross-section and increased absorptive losses, leading to a familiar trade-off between resonance visibility and \(Q\)-factor. Resonances with larger NPs are generally associated with lower \(Q\)-factors and higher peak extinctions. Furthermore, bigger particles generally induce a larger phase shift to the scattered light, leading to an increasingly redshifted resonance.

To validate our findings, we conducted 3D COMSOL simulations of the BRW system to reproduce the results obtained from the LSA. The simulations had an angle of incidence of 64 degrees and were performed for the BRW structure outlined in Table \ref{table:braggtech} with embedded periodic spherical gold NPs. We then subtracted the transmission spectra of the empty BRW from the obtained SLR spectra, allowing us to isolate the plasmonic resonance peak. Subsequently, the Fano function described in Eq.~\eqref{eq:fano} was fitted to estimate the resonance properties. The transmission, reflection, and fitted Fano function for selected simulations are shown in Fig.~\ref{fig:comsol}.

We note that the periodic boundary conditions in COMSOL simulations emulate infinitely large arrays. Corresponding calculations with the LSA are compared with the COMSOL simulations in Fig.~\ref{fig:QwithAuCOMSOL} for both changing particle radii and array period. While the two methods are qualitatively comparable, some simplifying assumptions in the LSA model make for differences in quantitative comparisons. Firstly, LSA does not account for the changing transmission through the Bragg walls. In COMSOL, diminished extinction is observed when SLR does not align perfectly with the transmission window, apparent in Figs. \ref{fig:comsol}(b) and (f). Secondly, as discussed at the end of Section \ref{ch:theory}, LSA does not take into account the uncertainty in the propagation wavenumber \(k_z\)~\cite{Hu:09}, which leads to LSA having systematically higher \(Q\)-factors as opposed to full wave analysis. Further discrepancies include the stronger redshift as a function of particle size in LSA compared to COMSOL. 

Unfortunately, the effect of array size on the \(Q\)-factor is unfeasible to investigate using COMSOL simulations. Corresponding multi-partite simulations would be computationally extremely heavy due to broadened simulation space. Nevertheless, the conducted simulations assure that the LSA method can yield qualitatively accurate results for the system.

\section{Conclusions}
Reducing the sizes of metasurfaces that exhibit SLRs is essential for their adoption in selected applications. We have both theoretically and numerically demonstrated a new approach to decrease array sizes of diffractive nanoarrays while retaining their high \(Q\)-factor values. Based on our results, the approach enables up to two orders of magnitude reductions in array dimensions, at the expense of increased structural thickness. Radiant \(Q\)-factors in the order of  10\(^4\) in arrays with dimensions smaller than 50\;\textmu m were achieved. This miniaturization of array area could be used in technologies, such as spatial light modulators, which demand pixel dimensions in the order of 10\;\textmu m. Scaling down plasmonic metasurface areas opens up possibilities also in applications like spectral imaging, where finer pixel sizes are advantageous. Furthermore, smaller arrays facilitate faster and simpler fabrication, especially with high-precision techniques like electron-beam lithography and focused ion beam milling, where writing areas are limited.

\section*{Acknowledgements}
We acknowledge the support of the Flagship of Photonics Research and Innovation (PREIN) funded by the Academy of Finland. JK also acknowledges the Magnus Ehrnrooth foundation for their PhD grant.

\printbibliography

\newpage
\newgeometry{left=2.54cm,right=2.54cm,top=25.4mm,bottom=2.54cm}
\onecolumn
\setcounter{section}{0}
\setcounter{figure}{0}
\setcounter{equation}{0}
\renewcommand{\theequation}{S.\arabic{equation}}
\renewcommand{\thefigure}{S.\arabic{figure}}
   \begin{center}
     {\huge\sffamily Supplemental material for Dispersion-induced \(Q\)-factor enhancement in waveguide-coupled surface lattice resonances}\\
   \end{center}
   
\section{Zeroth-order SLR}
Zeroth-order SLR is a special case of SLR for which the periodicity of the array has no effect on the wavelength of the SLR. They are a solution to
\begin{equation}{\label{eq:slrloc2}}
    k_{\mathrm{sub}} = k_\mathrm{inc}+mk_g\,,
\end{equation}
with \(m=0\). If the incidence is from air superstrate (\(n=1\)), the equation is reduced to 
\begin{equation}{\label{eq:zerothorderSLR}}
    n_{\mathrm{eff}}(\lambda) = \sin{\theta}.
\end{equation}
Equivalently, grating equation with \(m=0\) refers to Snell's law. However, in BRWs a resonance occurs when the refracted light matches with a waveguide mode. For normal incidence (\(\theta = 0\)) the resonance is at exactly \(n_{\mathrm{eff}}(\lambda) = 0\), and light is directly coupled to the epsilon-near-zero (ENZ) mode of the thin-films structure. Since constant phase over the propagation direction is observed, every possible NP embedded in the structure will automatically oscillate in-phase, adding to the resonance.

Tilting the incident angle, the zeroth-order SLR blueshifts with the ENZ/cavity mode as shown in Fig.~2(c). The situation is understood as the transverse incident wavenumber (\(k_{\mathrm{inc}}\)) matching the effective propagation wavenumber (\(k_z\)) of the waveguide mode.
\begin{equation}
    k_z = k_0 n_{\mathrm{eff}} = k_{z,\mathrm{inc}} = k_{\mathrm{0}}\sin{\theta}\,.
\end{equation}
Even with oblique angle of incidence, all NPs will be automatically in-phase with each other, due to incident transverse phase propagation matching the phase propagation in the substrate.

\begin{figure}[b!]
    \centering
    \includegraphics{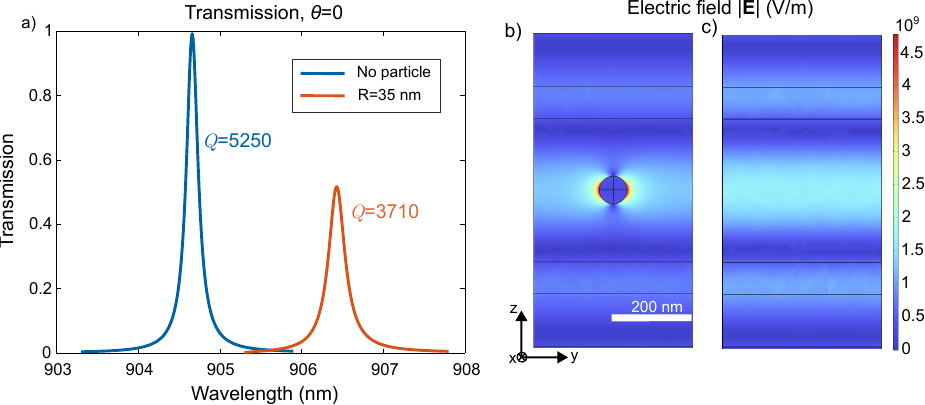}
    \caption{(a) Transmission spectra of ENZ mode and zeroth-order SLR from 3D COMSOL simulations. Electric fields at resonance for (b) zeroth-order SLR and (c) pure ENZ mode of the BRW structure.}
    \label{fig:zeroth}
\end{figure}

We investigated these zeroth-order resonances in COMSOL with normal incidence in BRWs with and without particles. The results are depicted in Fig.~\ref{fig:zeroth}. As expected, the pure ENZ mode had higher \(Q\)-factor, while the plasmonic particle increased local electric fields, and redshifted the resonance wavelength. 
\newpage

\section{Dispersion and uncertainty of wave numbers in leaky modes}
In a waveguide with non-perfect mirrors such as real metals or dielectric Bragg-reflectors losses will be introduced into the system as either radiative losses or material losses. We quantify these losses by making the transverse \(k_c\)-vector a complex quantity:
\begin{equation}{\label{eq:kccomplex}}
    k_c = k_c^{'} + \rmi k_c^{''}
\end{equation}
Now the propagation wavenumber \(k_z\) must fulfill leaky dispersion relation:
\begin{equation}{\label{eq:kzcomplex}}
    k^2 = k_z^2 + k_c^{2}.
\end{equation}
Equating the imaginary parts of the left and right-hand sides of the Eq.~\eqref{eq:kzcomplex} yields a relation between the imaginary (\(k_z^{''}\)) and real (\(k_z^{'}\)) parts of the propagation constant with the transverse components \cite{Cho2021}
\begin{equation}{\label{eq:dispprodderivation}}
\begin{aligned}
k^2 &= (k_z^{'}+\rmi k_z^{''})^2 + (k_c^{'}+\rmi k_c^{''})^2 \\
\mathrm{Im}[k^2] &= \mathrm{Im}[k_z^{'2}-k_z^{''2}+2\rmi k_z^{'}k_z^{''}+ k_c^{'2}-k_c^{''2}+2\rmi k_c^{'}k_c^{''}] \\
0 &= 2 k_z^{'}k_z^{''} +2 k_c^{'}k_c^{''}
\end{aligned}
\end{equation}
and consequently:
\begin{equation}{\label{eq:-kc}}
    k_z^{'}k_z^{''} = -k_c^{'}k_c^{''}
\end{equation}
Near the cutoff region \(k_c^{'} >> k_z^{'} \xrightarrow{} k_c^{''} << k_z^{''}\), and the wave vector of the mode is almost normal to the interfaces. Meanwhile, the fields are decaying (or increasing) extremely strongly in the propagation direction. 

Eq.~\eqref{eq:-kc} necessitates exponential growth in either transverse or propagation direction, a known problem of the leaky waveguide mode analysis. The problem is solved if the wavenumbers are considered to have a continuum of values as opposed to singular \(\delta\)-function-like values. As the waveguide mode assumes a Lorentzian distribution of transverse and propagation wavenumbers, the different background components become increasingly out-of-phase with each other. The different phases act as a nullifying effect for the exponential growth in the transverse direction. \cite{Hu:09}

\begin{figure}[b!]
    \centering
    \includegraphics{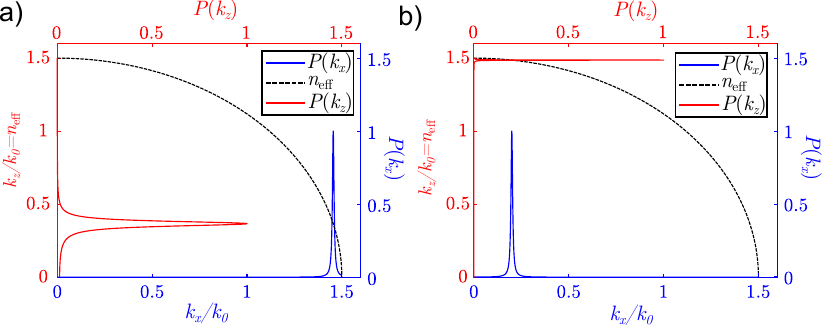}
    \caption{Lorentzian wavenumber distributions of a leaky waveguide mode (a) near the cut-off, and (b) far-away from the cut-off.}
    \label{fig:Lorentz}
\end{figure}

 Examples of the Lorentzian distributions are shown in Fig.~\ref{fig:Lorentz}, with \(\lambda = 800\)\;nm, \(\mathrm{Im}[k_c]=40\)\;mm\(^{-1}\), and \(n_\mathrm{core}=1.5\). Distributions \(P\) are normalized to one. Due to the relation~\eqref{eq:kzcomplex} between propagation and transverse wavenumbers, the Lorentzian linewidth for the effective index \(n_\mathrm{eff}\) is larger near the cut-off (Fig.~\ref{fig:Lorentz}a), than far-away from the cut-off (Fig.~\ref{fig:Lorentz}b). The half-width of the Lorentzian line shape is equal to the imaginary part of the wavenumber. 

Importantly, the Lorentzian distribution of propagation wavenumbers also affect the coupling between individual NPs inside the waveguide.  Consequently, the waveguide coupled SLRs have larger line widths than one would assume with a \(\delta\)-function-like mode wave vectors. Since the imaginary part of \(k_z\) is larger near the mode cut-off, the resonance broadening is expected to be greater near the cut-off. However, the effect is also expected to affect large arrays more, since coupling between far-away particles is affected more compared to neighboring particles. 

Including the effect in LSA would require a convolutional operation done for each term in the lattice sum, rendering Eq.~\eqref{eq:latticesum} to:
\begin{equation}{\label{eq:latticesumconvolution}}
    \dyad{S}(\lambda) = \omega^2\mu_0\sum_{j\neq i}^N \int_{-n_0k_0}^{n_0k_0} \dyad{G}_e(\vec{r}_i,\vec{r}_j,k_z)P_\lambda(k_z)\rmd k_z\,,
    \end{equation}
where \(P_\lambda(k_z)\) is the leaky-mode-specific wavenumber distribution at wavelength \(\lambda\). This modification however would render the computational complexity of the system unnecessarily high with minimal increase in the model accuracy. 

\section{Multi-partite unit cell LSA}
Multi-partite unit cell formulation for LSA is constructed in this chapter. The formulation is useful in lattices, such as honeycomb lattice, which can not be constructed with single-particle unit cell. The dipoles inside one unit cell may now have different dipole moments, but each unit cell is assumed identical as a whole. The effective polarizability:
\begin{equation}{\label{eq:eff}}
    \vec{p} = \dyad{\alpha}_{\textrm{eff}}\vec{E}_{\mathrm{inc}}\,,
\end{equation}
for a \(n\) -dipole unit cell system is now a \(3n\times3n\) block matrix with  \(\vec{p}\) and \(\vec{E}_{\mathrm{inc}}\) being \(3n\) sized vectors. The effective polarizability \(\dyad{\alpha}_\mathrm{eff}\) can be represented using the interaction dyadics \(\mathcal{G}_{l,k}\)~\cite{Kolkowski2020}:

\begin{figure}[b!]
    \centering
    \includegraphics[width=0.80\textwidth]{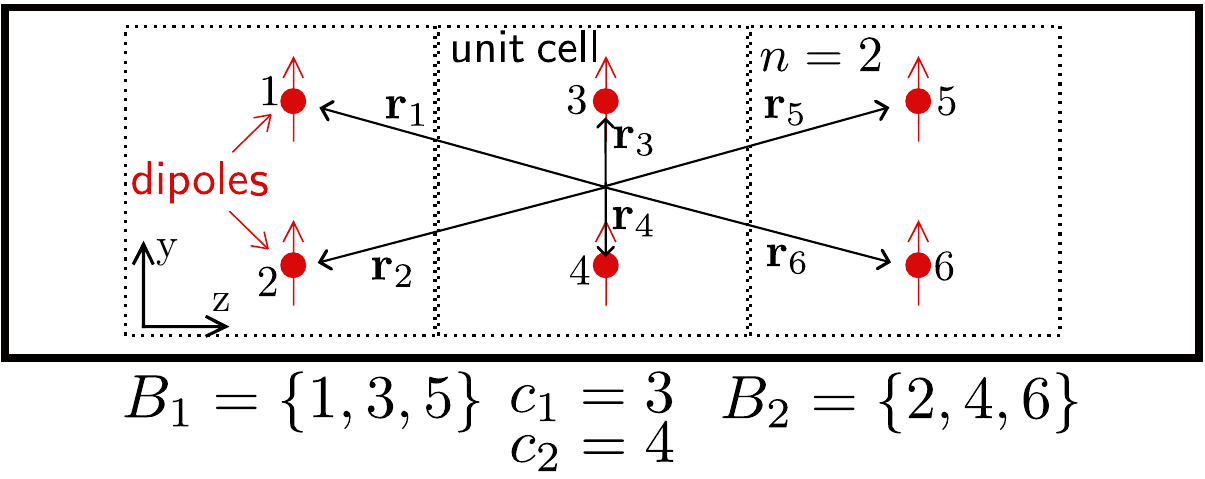}
    \caption{Schematic representation of LSA system of \(N=6\) dipoles with unit cell size \(n = 2\). The dipoles are divided into two groups \(B_1\) and \(B_2\) forming two sublattices. }
    \label{fig:schematicmath}
\end{figure}
\begin{equation}{\label{eq:eff_inv}}
\dyad{\alpha}_{\mathrm{eff}}=\text{inv}\left(\left[\begin{array}{cccccc}
\mathcal{G}_{1,1} & \mathcal{G}_{1,2} & \cdots & \mathcal{G}_{1, k} & \cdots & \mathcal{G}_{1, n} \\
\mathcal{G}_{2,1} & \mathcal{G}_{2,2} & \cdots & \mathcal{G}_{2, k} & \cdots & \mathcal{G}_{2, n} \\
\vdots & \vdots & \ddots & \vdots & & \vdots \\
\mathcal{G}_{l, 1} & \mathcal{G}_{l, 2} & \cdots & \mathcal{G}_{l, k} & \cdots & \mathcal{G}_{l, n} \\
\vdots & \vdots & & \vdots & \ddots & \vdots \\
\mathcal{G}_{n, 1} & \mathcal{G}_{n, 2} & \cdots & \mathcal{G}_{n, k} & \cdots & \mathcal{G}_{n,n}
\end{array}\right]\right)
\end{equation}
Each interaction dyadic \(\mathcal{G}_{l,k}\) represents how all particles in a sublattice \(k\) affect one (centrally located) particle in sublattice \(l\). Each interaction dyadic can be written as:
\begin{equation}{\label{eq:interactiondyad}}
    \dyad{\mathcal{G}}_{l,k}  = \sum_{j\in B_k} \dyad{A}_{c_l,j}\,,
\end{equation}
where \(B_k\) is a group of dipoles in sublattice \(k\), \(c_l\) is an index of a particle in the sublattice \(l\) and \( A_{c_l,j}\) determine the dipole interaction between particles \(i\) and \(j\), i.e.~\(\vec{E}_i = A_{i,j}\vec{p}_j\). \(A_{i,j}\) is written as:
\begin{equation}\label{eq:Aij}
\begin{aligned}
    A_{i\neq j} &= -\omega^2\mu_0 \dyad{G}_e(\vec{r}_i,\vec{r}_j) \\
     A_{ii} &= \dyad{\alpha}^{-1}
    \end{aligned}
\end{equation}
Using Eqs.~\eqref{eq:interactiondyad} and \eqref{eq:Aij}, the off- and on-diagonal interaction dyadics are written as:
\begin{equation}{\label{eq:interactiondyadneq}}
   \dyad{\mathcal{G}}_{l\neq k} = -\omega^2 \mu \sum_{j\in B_k}\dyad{G}_e(\vec{r}_{c_l}, \vec{r}_j ),
\end{equation}
\begin{equation}{\label{eq:dyadEqual}}
    \dyad{\mathcal{G}}_{l=k} = \dyad{\alpha}_{c_l}^{-1}-\omega^2\mu\sum_{j\in \{B_k \backslash c_l\}} \dyad{G}_e(\vec{r}_{c_l} ,\vec{r}_j).
\end{equation}
A schematic representation of a bipartite unit cell LSA system is presented in Fig.~\ref{fig:schematicmath}.

LSA reduces the computational complexity of conventional discrete dipole approximation \cite{YURKIN2007558} by summing over the Green's functions before the matrix inversion. Also, the reduced number of interactions taken into account in LSA decrease the computational complexity of the method. 

\section{Geometric series}
In this chapter, we derive the closed form of the interaction dyadics in multi-partite unit cell LSA for infinite number of lattice sites in a 1D PEC-waveguide with losses. We start the derivation by noting that the only \(z\)-dependence in the Green's function Eq.~\eqref{eq:Ge1} are the phase-propagation terms \(\rme^{\rmi k_zz}\) inside vector wave functions. Detaching the term from the functions, the Green's function is expressed as follows: 
\begin{equation}{\label{eq:Ge1_reduced}}
\begin{aligned}
\dyad{G}_{e 1}\left(\vec{R}, \vec{R}^{\prime}\right)
= & -\frac{1}{k^2} \hat{z} \hat{z} \delta\left(\vec{R}-\vec{R}^{\prime}\right) 
 +\frac{\rmi}{a b} \sum_{\mathrm{m}, \mathrm{n}} \dyad{G}_{\mathrm{m}\mathrm{n}} \rme^{\pm\rmi k_z(z-z^{\prime})}, z \gtrless z^{\prime} ,\\
\dyad{G}_{\mathrm{m}\mathrm{n}} &= \frac{2-\delta_0}{k_c^2 k_z}\left[\vec{M}_{\mathrm{m}\mathrm{n}}^*\left(\pm k_z\right) \vec{M}_{\mathrm{m}\mathrm{n}}^{*\prime}\left(\mp k_z\right)\right. 
  \left.+\vec{N}_{\mathrm{m}\mathrm{n}}^*\left(\pm k_z\right) \vec{N}_{\mathrm{m}\mathrm{n}}^{*\prime}\left(\mp k_z\right)\right]
\end{aligned}
\end{equation}
where \(^*\) notes that the vector wave function does not contain the phase-propagation term. Now taking into account the losses of the mode and incident angle, the interaction dyadic \eqref{eq:interactiondyadneq} for one mode-order (\(\mathrm{m},\mathrm{n}\)) comes into form
\begin{equation}{\label{eq:interactiondyadneqinfsum}}
   \dyad{\mathcal{G}}_{l\neq k, \mathrm{m}\mathrm{n}} = -\omega^2 \mu \sum_{j\in B_k}\dyad{G}_{mn}\rme^{\rmi(k_{\mathrm{inc},x}(x_{c_l}-x_j)+k_{\mathrm{inc},y}(y_{c_l}-y_j))}\rme^{\rmi (z_{c_l}-z_j)(\pm k_z+k_{\mathrm{inc},z}))}, z_{c_l} \gtrless z_j,
\end{equation}
where \(k_z\) is defined by Eq.~\eqref{eq:greenhelp}. In a 1D array, all particles in the same sublattice \(B_k\) have identical \(x\) and \(y\)-coordinates. Furthermore, all the unit cells are evenly spaced in \(z\)-direction with a period \(p\). The difference in \(z\)-coordinates can be expressed as
\begin{equation}
\begin{aligned}
    z_{c_l}-z_j = z_{c_l}-z_0-j^{\prime}p ,\\
    j\in B_k ,\text{ and } j^{\prime}\in \{-N^{\prime}/2,-N^{\prime}/2+1,\dots, N^{\prime}/2\},  
\end{aligned}
\end{equation}
where \(N^{\prime}\) is the total number of particles in sublattice \(B_k\). Eq.~\eqref{eq:interactiondyadneqinfsum} can be further simplified to 
\begin{equation}{\label{eq:interactiondyadneqinfsum2}}
\begin{aligned}
   \dyad{\mathcal{G}}_{l\neq k, mn} &= -\omega^2 \mu \dyad{G}_{mn}^{\prime}\left(1+\sum_{j^{\prime}=1}^{N^{\prime}/2}\left(\rme^{-\rmi p(k_z+k_{\mathrm{inc},z})j^{\prime}}+\rme^{-\rmi p(k_z-k_{\mathrm{inc},z})j^{\prime}}\right) \right)\\
   \dyad{G}_{mn}^{\prime} &= \dyad{G}_{mn}\rme^{\rmi (\vec{k}_{\mathrm{inc}}\cdot (\Vec{r}_{c_l}-\vec{r}_0) + (z_{c_l}-z_0)k_z )}
\end{aligned}
\end{equation}
Taking now \(N^{\prime}\xrightarrow{}\infty\), both terms inside the sum are revealed to be individual geometric series. Changing the series now to their closed forms finally grant us closed form of the interaction dyadic:
\begin{equation}
     \dyad{\mathcal{G}}_{l\neq k, mn} = -\omega^2 \mu \dyad{G}_{mn}^{\prime}\left(\frac{1}{1-\rme^{-\rmi p(k_z+k_{\mathrm{inc},z})}}+\frac{1}{1-\rme^{-\rmi p(k_z-k_{\mathrm{inc},z})}}-1 \right)\\
\end{equation}

Similarly the diagonal interaction dyadics can be expressed as 
\begin{equation}{\label{eq:closedformeqsupp}}
     \dyad{\mathcal{G}}_{l=k, mn} = \alpha^{-1}-\omega^2 \mu \dyad{G}_{mn}^{\prime}\left(\frac{1}{1-\rme^{-\rmi p(k_z+k_{\mathrm{inc},z})}}+\frac{1}{1-\rme^{-\rmi p(k_z-k_{\mathrm{inc},z})}}-1 \right)\\
\end{equation}
With the main difference arising from the self-interaction term \(\dyad{A}_{ii} = \dyad{\alpha}^{1}\), since \(z_{c_l} = z_0\), and the \(j^{\prime} =0 \) is referring to the same particle the interaction dyadic is calculated for.

\section{Fano fits for $Q$-factors}
\(Q\)-factors of different resonances were estimated by fitting Fano-resonaces to the obtained data. Fano-resonance function is given as~\cite{Wang:19}
\begin{equation}{\label{eq:fanosupp}}
    T_{\mathrm{fano}}(\lambda) = |a_1+a_2\rmi+\frac{I}{(\lambda-\lambda_0-\rmi\gamma)}|^2,
\end{equation}
where \(\lambda_0\) is the resonance wavelength, \(\gamma\) is the resonance half-width half-maximum, and \(a_1, a_2\) and \(I\) fitting parameters related to resonance shape and magnitude. \(Q\)-factors were obtained by the following relation:
\begin{equation}
    Q = \frac{\lambda_0}{2\gamma}
\end{equation}
To approximate the radiant \(Q\)-factor, fitting was done for both the imaginary and real parts of the lattice sum \(S\), separately. Then, a more reasonable fit was automatically chosen to increase robustness. Typically, \(Q\)-factors obtained from real and imaginary parts of the lattice sum deviated less than 1\% from each other.
\begin{figure}[b!]
    \centering
    \includegraphics[width=\textwidth]{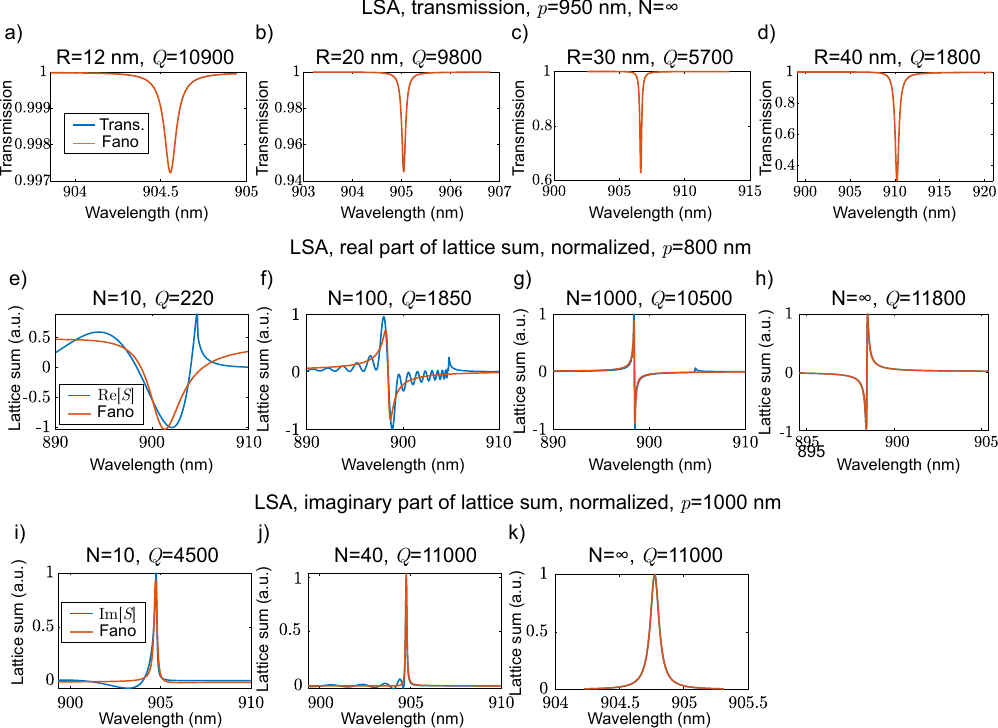}
    \caption{Fits to data. (a)--(d) Transmission with changing particle radii. (e)--(h) Real part of lattice sum. (i)--(k) Imaginary part of the lattice sum.}
    \label{fig:Fano}
\end{figure}
Fig.~\ref{fig:Fano} shows some of the used fits. Fig.~\ref{fig:Fano}(a)--(d) depicts Fano fits to transmission spectra obtained from LSA with \(N=\infty\), \(p=950\)\;nm, and different particle radii \(R\). Fig.~\ref{fig:Fano}(e)--(h) shows fits to the normalized real part of the lattice sum with different number of lattice sites with a period of \(p=800\)\;nm. Finally, Fig.~\ref{fig:Fano}(i)--(k) has fits for the normalized imaginary part of the lattice sum in a lattice with a period of \(p=1000\)\;nm. 

It is apparent that the fits for transmission, imaginary and real parts of the lattice sums are very good and reliable for infinite arrays. In finite systems, the resonance line shapes start deviating from the ideal Fano shape due to oscillations, caused by the LSA-assumption that all particles have identical dipole moments. Naturally, in finite systems, the particles near the edges of the array would experience weaker field enhancement and dipole moments, leading to the nonphysical oscillations in LSA. Nonetheless, the fits for finite systems are fairly reliable at approximating the \(Q\)-factor.

\section{Suppression of Modes in waveguide-SLRs}
\begin{figure}[b!]
    \centering
    \includegraphics[scale=0.8]{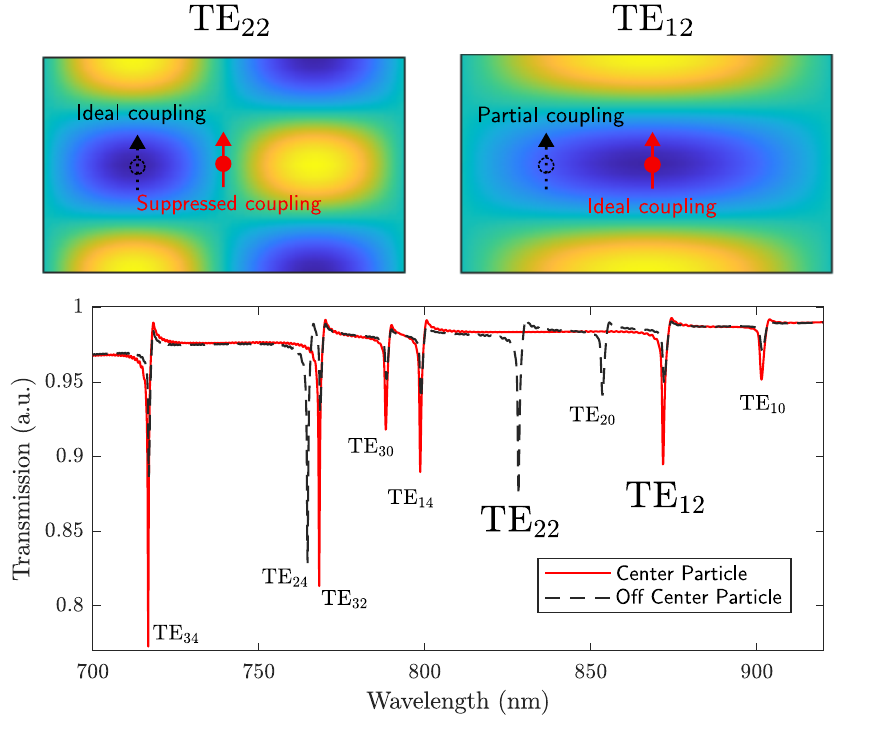}
    \caption{Suppression of different modes depending on NP position. a) Mode profile of TE\(_{22}\) -mode with two different particle locations. b) Mode profile of TE\(_{12}\) -mode with two different particle locations. c) Transmission spectra for both particle locations, showing multiple resonances arising from different waveguide-mode effective indices.}
    \label{fig:suppressionOfModes}
\end{figure}
Particle locations in waveguide-SLRs play a key role in determining the strength of the coupling between the particles~\cite{GERASIMOV2019}. This interplay can either heighten or suppress the coupling depending on particle locations in respect to the mode profile. In the dipole approximation, an SLR mode can be completely suppressed if all the particles reside in the nodes of the transverse mode profile of the coupling waveguide mode. Placing the NPs symmetrically in two antinodes of opposite phase also yield similar suppression. The effect is illustrated in Fig~\ref{fig:suppressionOfModes}. 

The excitation of a mode by a source inside the waveguide is proportional to \cite{jackson_1999Multipole_inwaveguide}
\begin{equation}
    \iiint \vec{J}\cdot \vec{E}_{\mathrm{m}\mathrm{n}}\rmd V
\end{equation}
where  \(\vec{J}\) is the electric current density of the source and \(\vec{E}_{\mathrm{m}\mathrm{n}}\) are the mode-fields. In a node of the field, the integral goes to zero as particle size vanishes, leading to no excitation. However, for particles with finite volume, higher-order terms in electric multipole expansion become important at describing the scattering. As such, the particle may excite a mode even from a node of the transverse profile. The treatment for multipole expansion of localized sources in PEC-waveguides is found in \cite{jackson_1999Multipole_inwaveguide}.

The effect of mode suppression could be used as an advanced waveguide grating for selective coupling of incident light into the waveguide. Allowing only one mode to couple, large multimode waveguides could function as single-mode waveguides while retaining their dimensions. Such devices might be used as a large-bandwidth single-mode waveguides. Due to the fixed array period, the devices would, however, necessitate the use of an original angle of incidence for each coupled wavelength.

\end{document}